\documentstyle[preprint,aps,psfig]{revtex}
\tightenlines
\begin{document}
\tighten
\draft
\preprint{
\vbox{
\hbox{IU/NTC 97-06}
\hbox{ADP--97--31/T264}
}}

\title{Testing Parton Charge Symmetry at HERA}
\normalsize
\author{J. T. Londergan}                                     
\address{
Department of Physics and Nuclear Theory Center, Indiana University,                                        \\
Bloomington IN 47408 USA                                  
}
\vspace{0.4cm}
\author{S. A. Braendler and A. W. Thomas}
\address{
Department of Physics and Mathematical Physics and \\ 
Special Research Centre for the Subatomic Structure of Matter, \\ 
University of Adelaide,                 
Adelaide S.A. 5005 Australia
}
\date{\today}
\maketitle
\begin{abstract}
There are strong theoretical indications that the minority valence
quark distributions in the nucleon may break charge symmetry by as
much as 3-5\%.  We show that a
comparison of $e^{\pm} D$ deep inelastic scattering through the charged
current, for example at HERA, could provide a direct test of this effect.
This measurement is also be sensitive to an intrinsic component of
the strange quark sea which leads to $s \neq \bar{s}$.
\end{abstract}
\hspace*{-0.5cm}

\vspace*{3cm}


\vfill\eject
The assumption of charge symmetry is an essential element of all
current, phenomenological analyses of deep inelastic scattering data
in terms of parton distributions.  We recall that charge symmetry is
the invariance of a Hamiltonian under a rotation by 180$^o$ about 
the 2-axis in isospace:                  
\begin{equation}
\left[ e^{i \pi I_2}, H_{CS} \right] = 0.
\label{eq:1}
\end{equation}
It is a much more restrictive symmetry than isospin (i.e., charge
independence), which requires $[I_i,H_{CS}] = 0, i=1,2,3$. 
(Here $I_i$ is the generator of the $i$'th component of rotations in
isospace.)  A charge symmetry transformation maps $n$ into $p$ and 
$d$ into $u$ (and vice versa), so charge symmetry at the quark
level requires
\begin{eqnarray}
d^n(x) &=& u^p(x) \hspace{2cm}; \hspace{2cm} u^n(x) = d^p(x), \nonumber \\
\bar{d}^n(x) &=& \bar{u}^p(x) \hspace{2cm}; 
\hspace{2cm} \bar{u}^n(x) = \bar{d}^p(x). 
\label{eq:2}
\end{eqnarray}
Charge symmetry thus reduces the number of light quark distributions 
to be extracted from data by a factor of two\cite{fec}. All parton
distributions for the neutron can be expressed in terms of those
for the proton.

Studies of charge symmetry and charge independence in nuclear systems
are very well developed\cite{mill}.  
While the latter is quite often broken at
the 5\% level, the former is usually reliable to 1\% or better.
Theoretical estimates of charge symmetry violation (CSV) in parton
distributions were only made a few years ago.  
This was motivated by the successful calculations of charge 
symmetric parton distributions in the nucleon\cite{adel,sst,str,bm}. 
In these calculations, parton distributions were obtained from
the relation
\begin{equation} 
q(x,\mu^2) = M \sum_X \left| \langle X \left| \psi_+(0)
\right| N \rangle \right|^2 \delta(M(1-x)- p_X^+) ~~.
\label{partcalc}
\end{equation}
In Eq.(\ref{partcalc}), $X$ represents a complete set of eigenstates
for the residual system. The parton distribution $q(x,\mu^2)$ is 
guaranteed to have proper support, i.e.\ it vanishes for $x>1$. 
One can easily see that the lowest-mass intermediate states dominate
the sum over states in the valence region and extensive investigation
has shown that simple models, like 
the MIT bag \cite{MIT}, are capable of 
explaining the shapes of the charge symmetric parton
distributions. 

Beginning with Sather\cite{sath} and Londergan 
{\it et al.}\cite{lond,lond1}, it was realized that the dominant 
contribution to parton CSV should arise from mass differences in
the eigenstates $p_X^+$ in Eq.(\ref{partcalc}), rather than from the
quark wavefunctions.  In this case, expressions for
the parton CSV terms could be obtained from the charge symmetric
parton distributions.  The dominant effect for valence quarks turned
out to be the $u-d$ mass difference for the spectators to the struck
quark.  As a consequence, the biggest percentage effect occurred in
the ``minority'' distribution $d^p(x) - u^n(x)$, where the $u-d$ mass 
difference enters twice (the valence spectators being (uu) and (dd), 
respectively).

Suppose we define the valence CSV quantities $\delta d_V$ and 
$\delta u_V$ as:
\begin{eqnarray}
\delta d_V(x) &=& d^p_V(x) - u_V^n(x)  ; \nonumber \\ 
\delta u_V(x) &=& u^p_V(x) - d_V^n(x),
\label{eq:3}
\end{eqnarray}
then it was found that $\delta d_V(x) \approx - \delta u_V(x)$\cite{lond1}.  
As the ratio $d_V/u_V$ becomes small at large Bjorken $x$\cite{doveru}, 
the relative CSV effect is
much bigger in $\delta d_V(x)/d_V(x)$ than $\delta u_V(x)/u_V(x)$. 
For example, Rodionov {\em et al.}\cite{lond} found that 
the latter was never greater
than 1\%, as expected, whereas the former could be as large as 5-10\% 
at intermediate values of $x$.  Sather\cite{sath} suggested 
that $\delta d_V(x)/d_V(x)$ could grow to 3\% -- a value
supported by recent work by Benesh and Goldman \cite{BG} -- c.f.
however, Ref.\cite{supp}.  

In summary, it seems very likely that the minority valence
distributions, $d^p_V(x)$ and $u^n_V(x)$, may break charge symmetry by 
3-5\%. Experimental confirmation of such an unexpectedly large effect
would be extremely important.  Until now the best proposal to
determine $\delta d_V(x)$ and $\delta u_V(x)$ has involved $\pi^{\pm} D$
Drell-Yan\cite{lond1}.  That analysis
is somewhat complicated by the possibility of CSV in the quark
distributions of the pion, that is a non-zero value of $\bar{d}^{\pi^+}
- \bar{u}^{\pi^-}$.
In this paper, we discuss the possibility of extracting information
on CSV by analyzing charged current deep-inelastic scattering from 
an isoscalar target. In particular, we examine the possibilities 
for comparing charged-current DIS from electrons or positrons on
deuterium, which could be performed in the future at HERA. 

At the enormous values of $Q^2$ that can be probed at HERA, 
charge current (CC) weak
interaction processes such as $e^- p \rightarrow \nu_e X$
are not impossibly suppressed with respect to the electromagnetic process
$e^- p \rightarrow e^- X$.  In future experiments at HERA, one should
have both $e^-$ and $e^+$ beams, and current plans call for 
colliding beams involving heavier 
nuclei, such as $D$, in a few years\cite{HERAfut}. The $(e^-,\nu_e)$
reaction proceeds through absorption of a $W^-$ by the target partons. 
It couples only to the positively charged partons in the target, so
that for a deuteron target the structure function (per nucleon) is
\footnote{We have also ignored any tiny
corrections which might arise because $s^p \neq s^n$ or $c^p \neq c^n$.}
\begin{equation}
F_1^{W^-D}(x) =  \left[ u^p(x) + \bar{d}^p(x) +u^n(x) + \bar{d}^n(x) 
  + 2 \bar{s}(x) +2 c(x) \right] / 2.
\label{Wminus}
\end{equation}
(For simplicity we denote the $(e^-,\nu_e)$ reaction by the
charge of the virtual $W$ absorbed by the target.)  Eq.(\ref{Wminus})
is true under the following conditions.  We assume that the measurements
occur at energies and momentum transfers well above heavy quark 
production thresholds.  We also neglect terms in the CKM quark 
mixing matrix of the order of $|V_{ub}|^2 \sim |V_{td}|^2 \sim
1\times 10^{-4}$. If we have a positron beam, then the 
$(e^+,\bar{\nu}_e)$ deep-inelastic reaction
measures only the negatively charged partons:
\begin{equation}
F_1^{W^+D}(x) =  \left[ d^p(x) + \bar{u}^p(x) +d^n(x) + \bar{u}^n(x) 
  + 2 s(x) +2 \bar{c}(x) \right] / 2.
\label{Wplus}
\end{equation}
Taking the difference of the $e^+$ and $e^-$ CC cross sections one therefore
has 
\begin{equation}
\delta{F_1}(x) \equiv F_1^{W^+D}(x) - F_1^{W^-D}(x) = 
  {\delta{d_V}(x) - \delta{u_V}(x)\over 2} + s(x) - \bar{s}(x) - 
  c(x) + \bar{c}(x).
\label{F1}
\end{equation}

Eq.(\ref{F1}) demonstrates that any difference between
these structure functions will arise from either parton CSV, or from
differences between the $s$ and $\bar{s}$ distributions (or 
$c$ and $\bar{c}$ distributions) in the nucleon.  Furthermore, 
the relation holds for all $x$ values.  
Since the charged current weak interaction process is rare,
it would be essential to have very accurate calibration of the
electron and positron flux.  Detector efficiencies would not be
a major problem, as the signal involves prominent jets on the
hadron side and large missing energy and momentum on the
lepton side.

If, as is commonly assumed, $s(x) = \bar{s}(x)$ and $c(x) = \bar{c}(x)$, 
Eq.(\ref{F1}) provides a direct measure of the CSV in the parton 
distributions.  On the other hand, there has been quite a lot of interest 
recently \cite{sdiff}-\cite{BM}
in the possibility (first discussed in Ref.\cite{ST}) 
that $s(x) - \bar{s}(x)$ might
be non-zero -- with at least some suggestion of experimental support for
the idea \cite{sdiffexp}. We show below the estimate of $s - \bar{s}$
calculated by Melnitchouk and Malheiro \cite{sdiff}. This is quite
dependent on the form factor at the NK$\Lambda$ vertex which is not well
known. Clearly it will be important to determine first 
whether or not $F_1^{W^+D} - F_1^{W^-D} ( \equiv \delta{F_1})$
is experimentally non-zero. The interpretation 
in terms of CSV, $s \neq \bar{s}$ or 
possibly both, can then be pursued in detail.

In order to illustrate the size and shape of the effect expected we
have evolved the values of $\delta{d_V}$ and  $\delta{u_V}$ 
calculated in Ref.\cite{lond} -- as
shown at $Q^2 = 10$ GeV$^2$ in Fig. 1(a) of Ref.\cite{lond1} -- to  
values of $Q^2$ appropriate to HERA. 
The NLO QCD evolution of the distributions was carried out using the
Mellin transformation 
technique. By the use of this technique one easily transforms the 
classical DGLAP equations\cite{fec,altpar} into a system of 
ordinary differential equations. This is principally done using Mellin
moments, 
\begin{equation}
M^{N}(Q^{2}) = \int _{0} ^{1} dx\, x^{N - 1} F (x,Q^{2})
\end{equation}
to transform from $x$-space to complex $N$-space. 
In order to regain the 
evolved distribution, once the evolution 
equations have been evaluated, the results must be transformed back to 
$x$-space by the inverse Mellin transform.
This is performed by a contour integral 
in the complex $N$-plane,
\begin{equation}
F(x,Q^{2}) = {1 \over \pi} \int _{0} ^{\infty} dz \mbox{ Im} 
\left[e^{i \phi} x^{-c -z e^{i \phi} } \: M^{n = c + z e^{i \phi}} (Q^{2}) 
 \right],
\label{eq:peter2}
\end{equation}
in which the contour of integration, and hence the value of $c$, must 
lie to the right of all singularities of $M^{N}$ in the complex $N$ 
plane. For all the practical calculations, we have chosen to use the 
same values for $z$ and $\phi$ 
that were used in the original paper of  Gl\"{u}ck, 
Reya and Vogt \cite{GRV}. 
The major advantage of using the Mellin 
transformation method for computing the evolution of the moments is that 
it does not involve the tremendous amount of computing time that 
previous methods employed. The resulting predictions for $\delta{u_V}$ and
$\delta{d_V}$ at $Q^2 =100, 400$ and $10000$ GeV$^2$ are shown in Fig. 1.
%
%

In order to indicate the relative size of the difference in the $e^{\pm}
D$ cross sections expected, we divide the difference  
$F_1^{W^+D} - F_1^{W^-D}$ by the average of $F_1^{W^+D}$ and 
$F_1^{W^-D}$ to obtain:
\begin{eqnarray}
R(x)  &\equiv& \frac{F_1^{W^+D}(x) - F_1^{W^-D}(x)}{F_1^{W^+D}(x) + 
  F_1^{W^-D}(x)}  \nonumber \\ 
  &=&  {\delta{d_V}(x) - \delta{u_V}(x) + 2(s(x) - \bar{s}(x)) \over 
  {\displaystyle\sum_{j=p,n}\left[u^j(x) + \bar{u}^j(x)+ d^j(x) 
  +\bar{d}^j(x)\right] 
   + 2(s(x) + \bar{s}(x))}} = 
  R_{CSV}(x) + R_s(x).
\label{RW}
\end{eqnarray}
In Eq.(\ref{RW}), $R_{CSV}$ is the charge-symmetry breaking 
term, and $R_s$ is the term proportional to $s-\bar{s}$.  In this
equation we have neglected terms proportional to $c$ and $\bar{c}$
since we expect them to be quite small, and in addition we know of
no quantitatively reliable model for this quantity -- see, however, 
Ref.\cite{MelT} for an estimate of the intrinsic charm of the proton in
the context of the HERA anomaly.
%
%

{}For the combination $F_1^{W^+D} + F_1^{W^-D}$ we evolved the CTEQ4Q
parton distributions \cite{CTEQ} from their starting scale ($Q_0^2 =
1.6$ GeV$^2$) to the same values of $Q^2$ used for $\delta{d_V}$ and 
$\delta{u_V}$. The results for $R_{CSV}$ are shown in Fig.\ 2, at several
values of $Q^2$. We see that the CSV contribution to the cross section
rises from 1\% at $x = 0.4$ to 2\% at $x = 0.6$.
The prediction is not shown above $x =0.8$ because 
the original bag model predictions were not reliable at very
large $x$ and the cross section is, in any case, too small in that
region. 

In view of the appearance of $s - \bar{s}$ in Eq.(\ref{F1}), we have  
taken the recent work of Melnitchouk and 
Malheiro \cite{sdiff} as an indication of the possible size of the
effect. These authors evaluated the strange quark
distributions in a ``meson-cloud'' picture, where the $\bar{s}$ 
sea arises from the kaon cloud
of the nucleon, while the $s$-sea arises from the spectator strange
baryon, namely $\Lambda$ or $\Sigma$. We have chosen to show their
prediction\cite{MMcalc} with a (monopole) form factor of mass 1 GeV, which is the
largest value consistent with the latest CCFR data \cite{sdiffexp}.
Figure 3 shows the corresponding ratio $R_s$ at the same values 
%
%
of $Q^2$ as shown in Figs.\ 1 and 2. Clearly the order of magnitude of
the effect caused by $s - \bar{s}$ is very similar to that arising from
CSV. Nevertheless, the shapes are completely different and one would
expect to be able to separate the two phenomena on the basis of the measured
$x$-dependence of $R$.

 In conclusion, we note that although 
both the CSV and strangeness effects are predicted to be
at the level of a few percent, the  measurement of the difference in
$e^{\pm} D$ cross sections at the momentum transfers typical of HERA 
is a very ``clean'' experiment. {\em Any deviation from zero, 
at any value of $x$, would be extremely interesting}, whether its origin
lies in CSV or intrinsic strangeness. Since there are both electron and
positron beams at HERA and future plans for deuteron beams,
we hope that this comparison can be made in the near future. 

{\bf Acknowledgements}

We would like to acknowledge helpful correspondence with W. Melnitchouk
and S. Wright. 
This work was supported by the Australian Research Council. One
of the authors [JTL] was supported in part by the US NSF under research 
contract NSF-PHY-9408843.  One author [JTL] would also like to thank
the Special Research Centre for the Subatomic Structure of Matter for 
its hospitality during the period this research was carried out.


%
\begin{figure}
\centering{ \begin{picture}(40,350)(250,0)
\psfig{figure=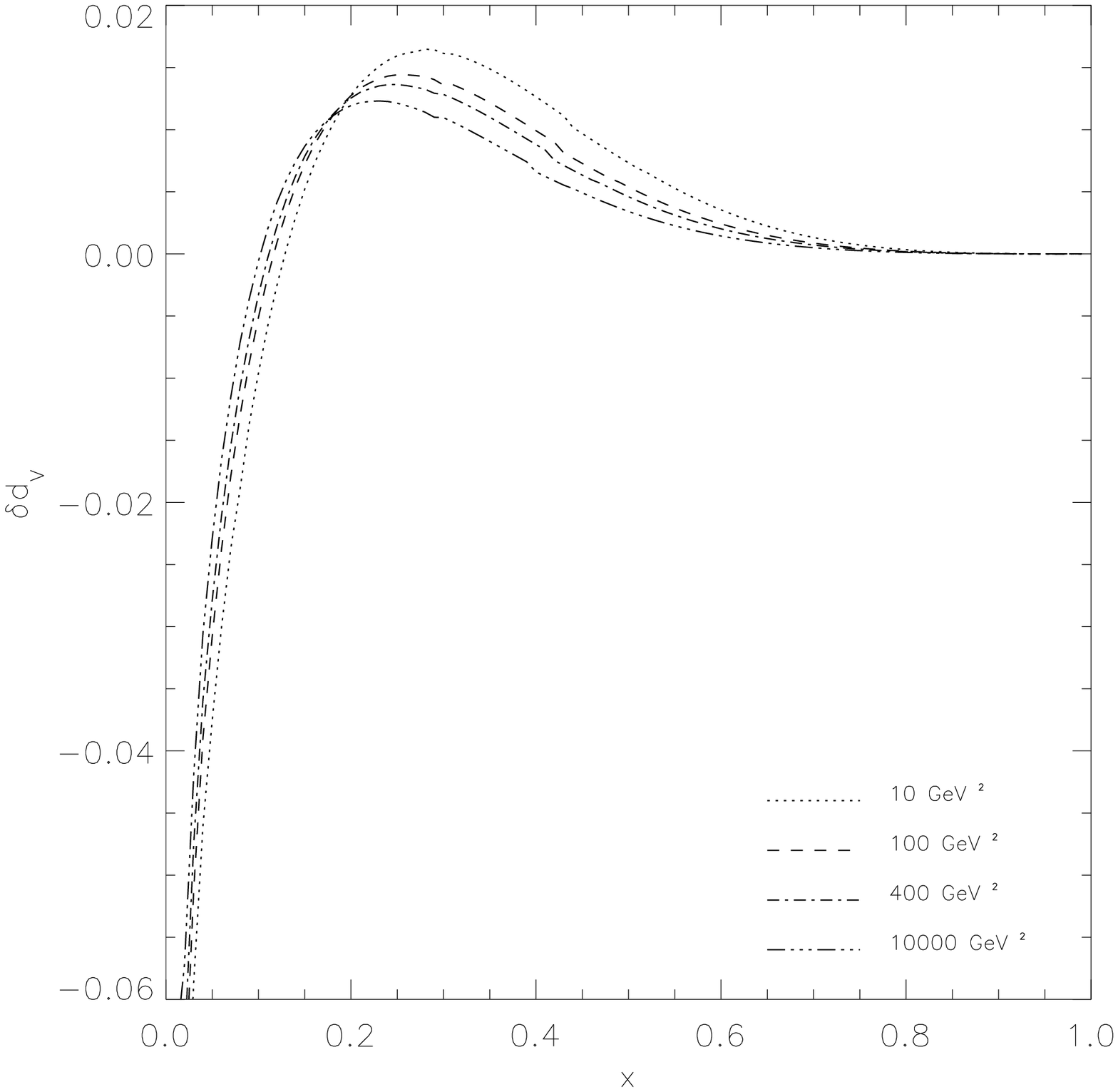,height=9cm}
\put(-10,0){\psfig{figure=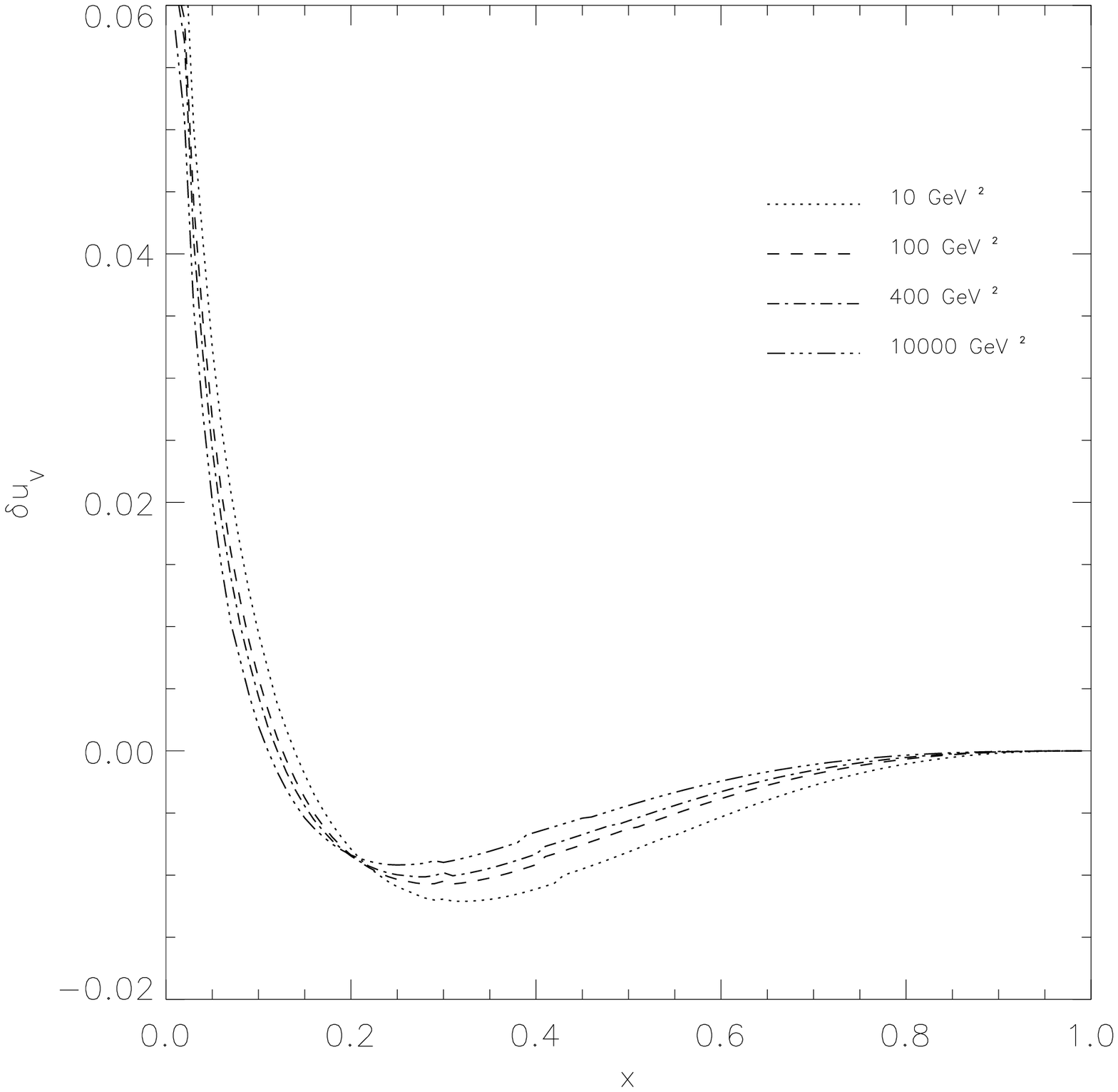,height=9cm}}
\end{picture}}
\caption{(a) The charge symmetry violation in the valence down quark
distribution.
(b) The charge symmetry violation in the valence up quark
distribution.}
\end{figure}
\newpage
\begin{figure}
\centering{\ \psfig{figure=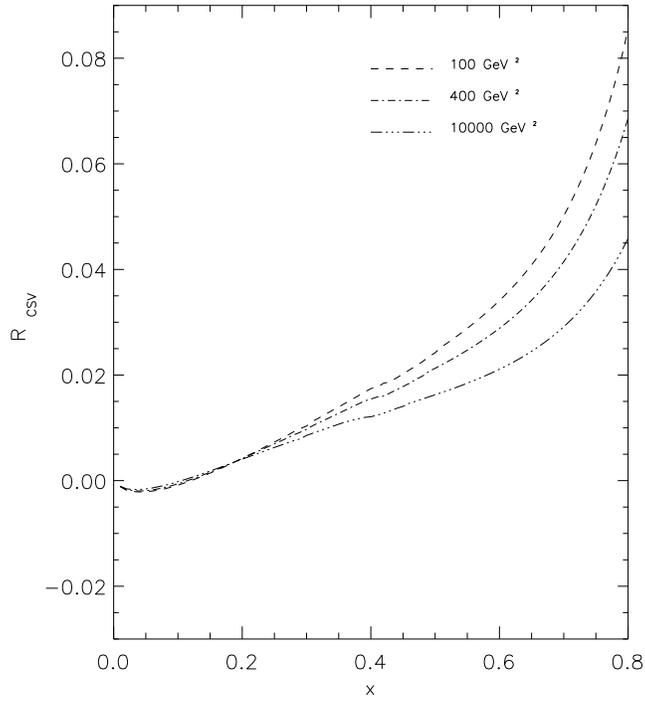,height=10cm}}
\caption{The charge symmetry violating ratio $R_{CSV}$ defined in
Eq.(10).}
\end{figure}
\begin{figure}
\centering{\ \psfig{figure=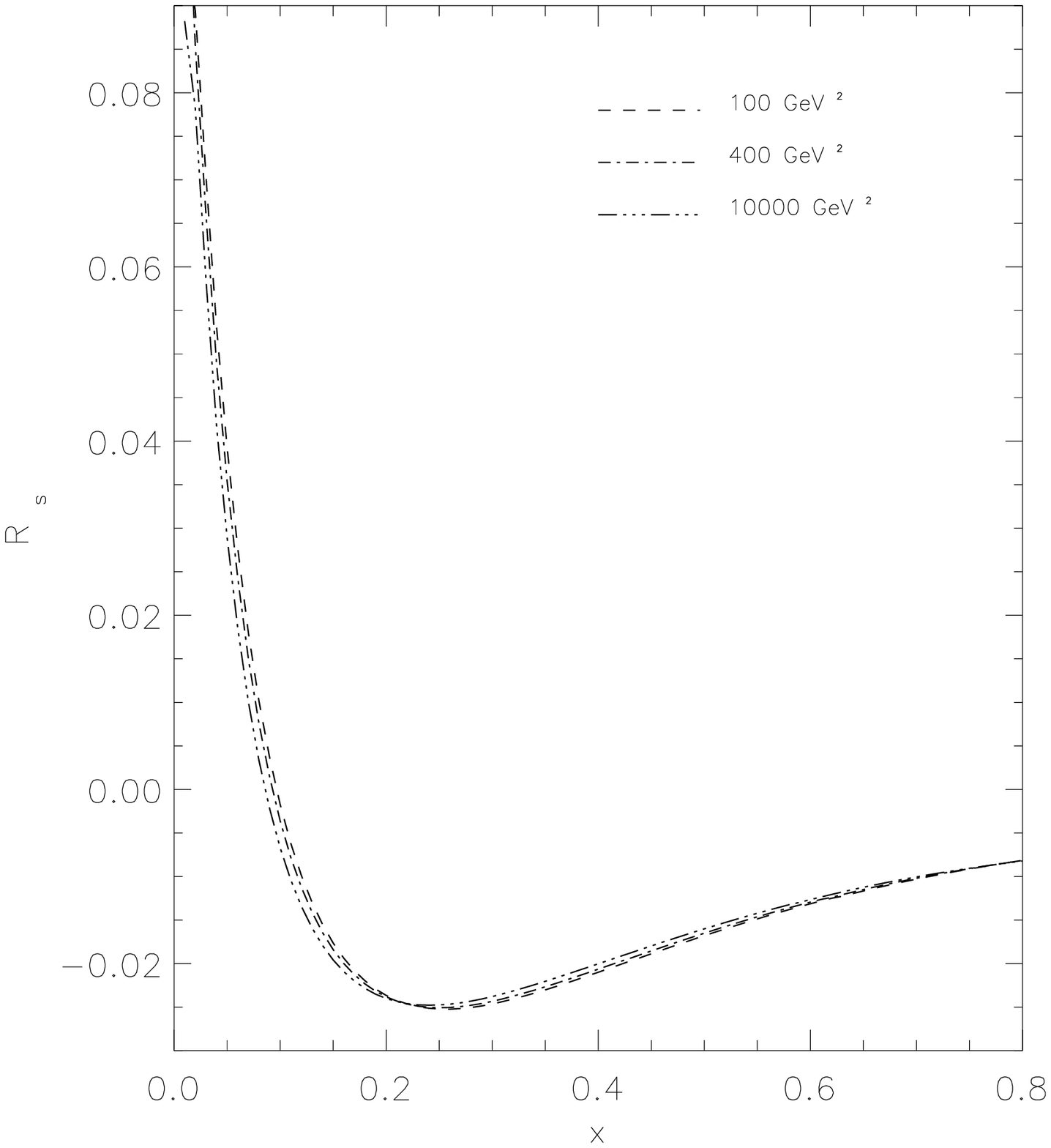,height=10cm}}
\caption{The relative contribution to the difference in $e^{\pm}$ cross
sections arising from a possible difference $s - \bar{s}$ -- labelled
$R_s$ in Eq.(10).}
\end{figure}

\begin{references}
\bibitem{MMcalc} We would like to thank W. Melnitchouk for sending us the
predictions of Ref.\cite{sdiff}.
\end{references}
\end{document}